\documentclass[useAMS,usenatbib,referee]{biom}

\def\bSig\mathbf{\Sigma}

\newcommand{\bO}{\boldsymbol{O}}

\newcommand{\bD}{\boldsymbol{D}}

\newcommand{\bZ}{\boldsymbol{Z}}

\newcommand{\bz}{\boldsymbol{z}}

\newcommand{\bbeta}{\boldsymbol{\beta}}
\newcommand{\bSigma}{\boldsymbol{\Sigma}}
\newcommand{\btheta}{\boldsymbol{\theta}}
\newcommand{\veta}{\boldsymbol{\eta}}

\def\P{ {\sf Pr}}

\newcommand{\cH}{{\mathcal H}}

\usepackage{outlines}
\usepackage{amsmath}
\usepackage{graphicx}
\usepackage[hidelinks]{hyperref}
\usepackage[dvipsnames]{xcolor}

\usepackage{mathabx}
\usepackage{threeparttable}
\usepackage{float}

\title[Time-In-Range Analyses with Inpatient Continuous Glucose Monitoring Data]{Time-In-Range Analyses of Functional Data Subject to Missing with Applications to Inpatient Continuous Glucose Monitoring}

\author{Qi Yu$^{1}$, 
Guillermo E. Umpierrez$^{2}$, and Limin Peng$^{1,*}$\email{lpeng@emory.edu} \\
$^{1}$Department of Biostatistics and Bioinformatics, Emory University, Atlanta,
Georgia, U.S.A.\\
$^{2}$Department of Medicine, Emory University, Atlanta, Georgia, U.S.A.}

\begin{document}

\pagerange{\pageref{firstpage}--\pageref{lastpage}} 

\label{firstpage}

\begin{abstract}
Continuous glucose monitoring (CGM) has been increasingly used in US hospitals for the care of patients with diabetes. Time in range (TIR), which measures the percent of time over a specified time window with glucose values within a target range, has served as a pivotal CGM-metric for assessing glycemic control.  However,  inpatient CGM is prone to a prevailing issue that a limited length of hospital stay can cause insufficient CGM sampling, leading to a scenario with functional data plagued by complex  missingness. Current analyses of inpatient CGM studies, however, ignore this issue and typically compute the TIR as the proportion of available CGM glucose values in range. As shown by simulation studies, this can result in considerably biased estimation and inference, largely owing to the nonstationary nature of inpatient CGM trajectories. In this work, we develop a rigorous statistical framework that confers valid inference on TIR in realistic inpatient CGM settings.  Our proposals utilize a novel probabilistic representation of TIR,  which enables leveraging the technique of inverse probability weighting and semiparametric survival modeling to obtain unbiased  estimators of mean TIR that properly account for incompletely observed CGM trajectories.  We establish desirable asymptotic properties of the proposed estimators.  Results from our numerical studies demonstrate good finite-sample performance of the proposed method as well as its advantages over existing approaches.  The proposed method is generally applicable to other functional data settings similar to CGM. 
\end{abstract}


\maketitle

\section{Introduction}
\label{s:intro}
Continuous glucose monitoring (CGM), which measures interstitial glucose every 1-5 minutes, can provide a panoramic view of  glycemic profile, thereby facilitating insulin adjustments and helping prevent
hypoglycemia in patients with diabetes \citep{spanakis2022continuous}.   With the capacity of wirelessly transferring glucose data, the CGM  telemetry  system  also enables remote glucose management that can reduce labor and potential infectious exposures of health care professionals.  With these benefits, the  use of CGM in US hospitals is on a rapid rise since  the COVID-19 pandemic. 

Time in range (TIR),  which measures the percentage of time that  glucose readings are within a target glycemic range (e.g., 70 - 180 mg/dL) over a specified amount of time  (e.g., 7 days), has served as a key metric for the assessment of glycemic control based on CGM \citep{diaTribe2022}.  TIR analyses,  which generally compare mean TIR between/among groups,  have been frequently adopted to evaluate the glycemic effect of an intervention or exposure of interest \citep{capaldo2020blood, linden2021real, mayeda2020glucose, al2022frequency, spanakis2022continuous, castaneda2022predictors, raj2022time}. 
However, existing TIR analyses often overlook a prevailing issue of inpatient CGM studies. That is, variable CGM sampling durations  across individuals, as limited by different lengths of hospital stay,  can result in CGM glucose trajectories that are not completely observed for the required amount of time. The common practice is to estimate the mean TIR by averaging the observed percentage of recorded CGM glucoses within the target glycemic range across individuals \citep[e.g.,]{cuesta2018characterization, olsen2024statistical, gecili2021functional}. As shown by our simulation studies which closely mimic real inpatient CGM data scenarios,  this approach, which simply ignores all missing CGM glucose readings,  can result in   biased estimation of mean TIR and misleading conclusions on group comparisons (see Table \ref{tab:simulation}).

With CGM glucose trajectories naturally formulated as functional data, the main statistical problem pertaining to the TIR analysis of an inpatient CGM study is how to estimate and infer about  the mean TIR with functional data subject to missing. Of note, different missing data patterns are present in the inpatient CGM setting. 
For an individual with CGM sampling not meeting the required length due to a short hospital stay, 
 the glucose  trajectory suffers from monotone missing  after the hospital discharge.  For all individuals, including those who stay in the hospital long enough, intermittent missing of CGM readings may occur due to various pragmatic reasons, such as random device errors or  temporary device removal required for performing other medical tests or procedures. Another remarkable statistical complication is that the missing mechanism underlying the inpatient CGM data may not be completely at random. It is intuitive to believe that better glycemic control (e.g. higher TIR for the range 70-180 mg/dl) may be associated with shorter hospital stays, which are closely related to the occurrence of the monotone missing. In addition, the glucose excursion patterns of hospitalized patients are typically non-stationary over time. As illustrated in Figure \ref{fig:2}(B), the glucose levels of hospitalized patients tend to be higher  in the first a few days during the hospital stay, and then stabilize as the glucose control improves under the hospital care. Thus, it is implausible to assume that the incompletely observed CGM glucose trajectories truncated by hospital discharges would adequately represent the underlying complete CGM trajectories.

Despite increasing attention to the missing data problem  in CGM studies, very few strategies are formally developed for  tackling such a problem. A conservative proposal is to exclude CGM data collected on days with over 20\% of CGM readings missing \citep{wilson2023cgm}.  This approach, however, does not takes into account the non-stationary  trend of inpatient glucose trajectories  and would completely neglect the monotone missing such as that resulted from a short hospital stay in inpatient CGM studies. \citet{klarskov2022telemetric} and \citet{olsen2022glycemic} proposed to employ latent Gaussian processes to impute missing data. While this approach is capable of handling the intermittent missing in CGM glucose trajectories, it has limitations in terms of accounting for the monotone missing.  To the best of our knowledge,  little effort was devoted to properly address  the important realistic missing data features inherent to inpatient CGM studies or other similar applications in order to provide statistically sound TIR analyses.

In this work, we fill in this gap by developing a rigorous statistical framework tailored to conduct TIR analyses with functional data  subject to mixed types of missing which may not be completely at random. We first discover a useful probabilistic representation of mean TIR. This representation allows us to  leverage the technique of inverse probability weighting, coupled with flexible semiparametric survival modeling, to derive unbiased estimates for mean TIR that simultaneously account for both intermittent missing and monotone missing under reasonable missing-at-random assumptions. By the proposed estimation strategy, we circumvent data imputation, which usually requires stronger assumptions for justification.
We establish the asymptotic properties of the proposed estimators and develop inference procedures including variance estimation and comparing mean TIR between/among groups.  The proposed estimation and inference procedures confer a much-needed analytical tool for performing rigorous TIR analyses of inpatient CGM data, which is  a timely effort in response to the rapid increase in hospital use of CGM. The new tool can readily be adapted to similar application settings, for example, inpatient use of other continuous monitoring systems, such as cardiac monitoring \citep{walsh2014novel} and body temperature monitoring \citep{smarr2020feasibility}. 

The rest of this paper is organized as follows. In Section \ref{sec:method}, we present the new representation of mean TIR and the proposed estimators of mean TIR, along with the adopted assumptions. In Section \ref{sec:inference}, we establish the asymptotic properties of the proposed estimators and discuss  inference procedures. In Section \ref{sec:simulation}, we report comprehensive simulation studies, which evaluate the finite-sample performance of the proposed method. An application to a recent inpatient CGM study is presented in Section \ref{sec:realdata}. The paper concludes with several remarks  in Section \ref{sec:remarks}.

\section{The Proposed Method}
\label{sec:method}
\subsection{Data and notation}
\label{sec:Method-background}  
Time in range (TIR)  is defined as the percentage of time that CGM glucose values are within a target glycemic range, denoted by $G$, over a specified amount of time, represented by the time interval $[0, \tau]$.
For example, to represent the TIR over 7 days with the target glucose range between 70 and 180 mg/dL, we set $G=[70, 180]$ and $\tau=7$ (days).  Let $\mathcal{T} \doteq \left\{ 0 = t_{1} < t _{2} < \dots < t_{K} = \tau \right\}$ denote the set of time points when the CGM sensor reads interstitial glucose over the time interval $[0, \tau]$ (e.g., a time grid equally spaced by 5 minutes between time $0$ and $\tau$). We formulate an individual's CGM glucose trajectory $Y(\cdot)$ as a right continuous piecewise constant function that only jumps at the time points in $\mathcal{T}$.  With notation defined above,  the TIR  can be represented as a statistical functional of $Y(\cdot)$:
\begin{equation}
    W = \frac{\int _{0} ^{\tau} I \left( Y(t) \in G \right) dt }{\tau}.
    \label{eq:TIR}
\end{equation}

In a CGM study conducted in hospital,  insufficient CGM sampling often occurs  due to a short hospital stay. Let $C$ denote the follow-up duration of CGM (i.e., time from the beginning to the end of CGM). If $C<\tau$, then $Y(t)$ with $t>C$ is not observed, resulting in monotone missing of the CGM trajectory.   Let $\delta^*(t)$ be a right continuous piecewise constant function. We use $\delta^*(t)$ to account for the intermittent missing of the CGM trajectory due to reasons such as random device errors. Specifically, for $t<C$, we have $\delta^*(t)=1$ when a CGM reading is available in the time interval $\left[ t_{\iota (t)}, t_{\iota (t) + 1} \right)$, where $\iota (t) = \text{max} \{ j: t_j \leq t < C, 1\leq j \leq K \}$ and 0 otherwise. Define $\delta(t)=\delta^*(t)I(t\leq C)$.

In addition to the CGM glucose trajectory $Y(\cdot)$, other information may also be observed or collected in an inpatient CGM study, for example, individuals' demographics and clinical characteristics. We capture such information by a $p$-dimensional  vector of covariates, $\bZ(\cdot)$, which is allowed to change over time.  The observed data then include $n$ independent and identically distributed (i.i.d.) replicates of 
$
    \left\{ \delta (t), C, \delta (t) Y(t), \boldsymbol{Z}(t): t \in ( 0, \tau ] \right\}$, 
denoted by $\left\{ \delta _i (t), C _i, \delta _i (t) Y_i (t), \boldsymbol{Z_{i}}(t): t \in [0, \tau ]\right\}_{i=1}^n$.

\subsection{Estimation of Mean TIR}
\label{sec:Method-Estimation}

Mean TIR, namely $\mu_W\doteq E(W)$, is often used to summarize TIR at the population level and has been commonly used as a primary endpoint to evaluate group differences in CGM studies. 
When all CGM glucose trajectories (i.e., $Y_i(\cdot),\ i=1,\ldots, n$) are completely observed over the time interval $[0, \tau]$, the mean TIR, $\mu_W$,  can be readily estimated by its empirical counterpart, namely,
$n^{-1}\sum_{i=1}^n \left(\int_0^\tau I(Y_i(t)\in G)dt\right)/\tau$. 

In practice, some $Y_i(\cdot)$'s are only partially observed, as often encountered in an inpatient CGM study. In the routine TIR analysis, one may estimate $\mu_W$  by simply averaging the within-individual proportion of the observed CGM readings belonging to $G$. This is equivalent to adopting a naive estimator of $\mu_W$, which is given by 
$$\widecheck\mu_W= n^{-1}\sum_{i=1}^n \widecheck W_i,$$ 
where 
$\widecheck W_i=\{\int_0^\tau  I(Y_i(t)\in G)\delta_i(t) dt\}/\{\int_0^\tau \delta_i(t) dt\} $. 
 However, the glucose trajectory of a hospitalized patient  is usually not stationary over time (see examples in Figure \ref{fig:2}(A)). As a result, $\widecheck W_i$, which is obtained from a ``snap shot'' over a shorter time period with $\delta_i(t)=1$, may  fail to adequately reflect $W_i$, the underlying TIR of individual $i$. Intuitively, this would lead to biased estimation of $\mu_W$. 
 This is confirmed by our simulation studies (see Table \ref{tab:simulation}).

To address the issue caused by the incompletely  observed $Y_i(\cdot)$'s, we propose an estimator of $\mu _{W}$ by discovering the following  probabilistic representation of the mean TIR: 
\begin{align}
    \mu _{W}  &= E \left( \frac{\int _{0} ^{\tau} I \left( Y(t) \in G \right) dt }{\tau} \right) = \frac{\int _{0} ^{\tau} p_{G} (t) dt }{\tau},
    \label{eq:muw_proposed}
\end{align}
where $p_{G} (t) =\Pr \left( Y(t) \in G \right) $, representing the probability of CGM glucose falling into the target range at time $t$. This representation of $\mu_W$ enlightens a viable direction to estimate $\mu_W$ based on an estimate for $p_G(t)$.  Adopting this strategy can lead to multiple benefits: (i) $p_G(t)$ is a population quantity depending on $Y(t)$ cross-sectionally; thus tackling $p_G(t)$ (instead of $W_i$) allows for information sharing across subjects and circumventing the difficulty in directly imputing  the missing data in  within-subject  glucose trajectories; (ii) with a proper estimate for $p_G(t)$ bounded between $0$ and $1$, the corresponding plug-in estimate for $\mu_W$ automatically satisfies the bounded constraint for $\mu_W$. 

To estimate $p_G(t)$, we need to properly account for the possible missingness of  $Y(t)$'s at time $t$. To tackle this problem, we adopt the following assumptions:

(M1): $\delta^*(t)$ is independent of $C$ and $Y(t)$; 

(M2):  $C$ is independent of  $Y(t)$ conditional on $\cH(t)$,  where $\cH(t)$ denotes the subject-specific history data before time $t$. 

Assumption (M1) is reasonable in CGM studies because the intermittent missing is mostly caused by some external technical issues with the CGM device  and is less likely related to the underlying glycemic control.  
Assumption (M2) admits  a realistic missing-at-random (MAR) mechanism for the monotone missingness related to $I(t\leq C)$. By this assumption,  $I(t\leq C)$ is independent of the current glucose value $Y(t)$ conditional on $\cH(t)$, while $I(t\leq C)$ is allowed to depend on the glucose history captured by $\cH(t)$, thereby accommodating realistic scenarios, such as  better glycemic control leading to a shorter hospital stay. 

We  can derive an estimator  of $p_G(t)$ by applying the idea of inverse probability weighting. Specifically, under assumptions (M1) and (M2), we can show that 
\begin{eqnarray*}
&&E \left\{ \frac{\delta^*(t) I(t\leq C) I(Y (t) \in G)}{\Pr(t\leq C | \mathcal{H}(t)) }\right\} 
=E \left[E\left\{\frac{\delta^*(t) I(t\leq C) I(Y (t) \in G)}{\Pr(t\leq C | \mathcal{H}(t))}\big |\cH(t)\right\} \right]\\
&&\hskip 0.2in
=E \left[\frac{E\{ \delta^*(t) I(Y (t) \in G)|\cH(t)\}\Pr(t\leq C | \mathcal{H}(t))}{\Pr(t\leq C| \mathcal{H}(t)) }\right]
=E\{ \delta^*(t) I(Y (t) \in G)\}= E\{\delta^*(t)\} p_G(t).
\end{eqnarray*}
Similarly, we can get
$
E \left\{ \frac{\delta^*(t) I(t\leq C) }{\Pr(t\leq C | \mathcal{H}(t)) }\right\} =E\{\delta^*(t)\}.
$
By these results, we propose the following  estimator of $p_G(t)$:
\begin{equation}
    \widehat{p} _{G} (t) = 
    \frac{\sum _{i = 1} ^{n} \delta_i(t)
    I(Y_{i} (t) \in G)/{\widehat{p}_{C, i}(t) }}
    {\sum _{i = 1} ^{n} \delta_i(t) /\widehat{p}_{C,i}(t)}
    \label{eq:ptilde}
\end{equation}
where $\widehat p_{C, i}(t)$ is a reasonable estimator of $p_{C, i}(t)\doteq\Pr(t\leq C_i|\cH_i(t))$ and $\cH_i(t)$ is the sample analogue of $\cH(t)$ (i.e., the history data of subject $i$ at time $t$).

{\it Remark 1}: Consider a special case where $C$ is independent of both $Y(t)$ and $\cH(t)$ for all $t\in[0, \tau]$. This points to a practical scenario where the termination of CGM is completely independent of the underlying CGM glucose trajectory. In this special case, the proposed estimator reduces to a simplified estimator,
$$\widetilde p_G (t)=    \frac{\sum _{i = 1} ^{n} \delta_i(t)
    I(Y_{i} (t) \in G)}{\sum _{i = 1} ^{n} \delta_i(t)}.
    $$ 

The remaining task is to obtain $\widehat p_{C, i}(t)$. To this end, we assume that the distribution of $C$ is influenced by $\cH(t)$ through the covariates in $\bZ(t)$, which may include summaries of glucose history up to time $t$ (e.g., average or  maximum CGM glucose before time $t$).  Following this direction, we further  model the relationship between $C$ and $\bZ(t)$. A popular choice of such a model is the semiparametric Cox proportional hazard model, which assumes
\begin{equation}
    \lambda _{C} (t | \mathcal{H}(t))
    = \lambda_{0}(t)\text{exp} \left\{ \boldsymbol{Z}(t)^{\top} \boldsymbol{\beta}_0 \right\},
    \label{eq:cox}
\end{equation}
where  $\lambda _{C} (t | \mathcal{H}(t))$ denotes the conditional hazard function of $C$ given $\cH(t)$, $\lambda_{0}(t)$ denotes an unspecified baseline hazard function, and $\boldsymbol{\beta}_0$ is a vector of unknown coefficients.  Let $\Lambda_0(t)=\int_0^t \lambda(u)du$. Using the observed data on $C$ and $\bZ(t)$, we can obtain the partial likelihood estimator of $\bbeta_0$, denoted by $\widehat\bbeta$, and the Breslow estimator of $\Lambda_0(t)$, denoted by $\widehat\Lambda(t)$ \citep{andersen1982cox, Lin2007}. The corresponding estimator of $p_{C, i}(t)$ is then given by 
\begin{equation}
\widehat p_{C, i}(t)= \exp\left\{-\int_0^t \exp\{\bZ_i(u)^{\top} \widehat\bbeta\}d\widehat\Lambda(u)\right\}.
       \label{eq:est_propensity}
\end{equation}

Plugging the $\widehat p_G(t)$ obtained based on \eqref{eq:ptilde} and \eqref{eq:est_propensity} into the representation of $\mu_W$ in \eqref{eq:muw_proposed} leads to the proposed estimator of $\mu_W$, 
\begin{equation}
\widehat\mu_W=\left\{ {\int _{0} ^{\tau} \widehat p_{G} (t) dt }\right\}/\tau.
\label{eq:proposed_est}
\end{equation}
Note that $\widehat p_G(t)$ is a piecewise constant function. Thus, the integration involved in  $\widehat\mu_W$ can be evaluated via finite summation. 

The proposed estimation of $\mu_W$ can readily accommodate other estimators of $p_{C, i}(t)$ obtained under different modeling of $C_i$ given $\cH_i(t)$, such as the additive hazards model \citep{Aalen1980}.  More specifically, we may consider a general setting where 
\begin{equation}
\Pr(t\leq C|\cH(t))=
\Psi(\veta(\bZ, t), t|\btheta_0).
\label{general_model}
\end{equation}
Here $\veta(\bZ, t)=\{\bZ(u):\ 0\leq u \leq t\}$ and $\btheta_0$ captures either real valued or function valued model parameters. It is easy to see that the Cox model \eqref{eq:cox} is a special case with $\Psi(\veta(\bZ, t), t|\btheta_0)=\exp\left\{-\int_0^t \exp\{\bZ(u)^{\top}\bbeta_0\}d\Lambda_0(u)\right\}$. 
The estimator of $p_{C, i}(t)$ can be generally  formulated as $\Psi(\bar\bZ_i(t), t|\widehat\btheta)$, where $\widehat\btheta$ is an existing estimator of $\btheta_0$, 

\section{Asymptotic Theory and Inference}
\label{sec:inference}

We study the asymptotic properties of the proposed estimator $\widehat\mu_W$.  Define ${\cal Z}(M)$=\{$f(t): [0, \infty)\rightarrow R^p$: each component of $f(\cdot)$ has a bounded total variation for $t\in [0, \tau]$ and $\sup_{t\in[0, \tau]}\|f(t)\|<M$\}. We assume the following regularity conditions:
\begin{outline}
\1[] (C1) There exists a positive constant $M$ such that $\bZ_i(\cdot)\in {\cal Z}(M)$ for $i=1,\ldots, n$;
\1[] (C2) (i) $\inf_{t\in[0, \tau]} E\{\delta(t)\}>0$; 
(ii) $\inf_{t\in[0, \tau]}  \Pr(t\leq C|\cH(t)\}>\nu$ for some constant $\nu>0$.
\end{outline}
Condition (C1) assumes bounded covariates which are often met in practice. Condition (C2) ensures positive probabilities for the  availability of $Y(t)$ throughout the time interval $[0, \tau]$, which is crucial for the identifiability of mean TIR. 

In addition, we suppose that the estimation of model \eqref{general_model}  can be carried out with existing methods, as exemplified for the special case of the Cox model \eqref{eq:cox}. We  assume the following condition to ensure the resulting estimator of $p_{C, i}(t)$ is adequate.

\begin{outline}
\1 [] (C3) There exists a Donsker 
function class \citep{vaart2023empirical}, 
$\{\xi(\bz, t, \cdot): \bz\in{\cal Z}(M),\ t\in[0, \tau]\}$, such that 
$$
\sup_{\bz\in{\cal Z}(M),\ t\in[0, \tau]} |n^{1/2} \{\Psi(\veta(\bz, t), t|\widehat\btheta)-\Psi(\veta(\bz, t), t|\btheta_0)\}- n^{-1/2} \sum_{i=1}^n \xi(\bz, t, \bO_i)|\rightarrow_p 0,
$$ where 
 $E\{\xi(\bz, t, \bO_i)\}=0$ and
$\bO_i$ denotes the observed data for subject $i$. 
\end{outline}
By  the results of \cite{Lin2007} and \cite{Peng2007regression},  the $\widehat p_{C, i}(t)$ given in \eqref{eq:est_propensity} satisfies condition (C3) under mild assumptions. Condition (C3) is also expected to hold  when  many other semiparametric regression models for $C_i$ given $\cH_i(t)$ are adopted to estimate $p_{C, i}(t)$.

Under regularity conditions (C1)--(C3), we first establish the asymptotic properties of $\widehat p_G(t)$ and then show that $\widehat\mu_W$ is consistent and asymptotic normal. These results are stated in the following two theorems.

{\it Theorem 1. Under regularity conditions (C1)--(C3), we have $\sup_{t\in[0, \tau] } |\widehat p_G(t)-p_G(t)|\rightarrow_p 0$ and $n^{1/2} \{\widehat p_G(t)-p_G(t)\}$ converges weakly to a mean zero Gaussian process with covariance $E\{\phi_1(t)\phi_1(s)^{\top}\}$, where $t$ and $s\in[0, \tau]$ and $\phi_1(t)$ is defined in equation {(S13)} of Web Appendix A.}

{\it Theorem 2. Under regularity conditions (C1)--(C3), we have $\widehat\mu_W\rightarrow_p \mu_W$ and $n^{1/2} (\widehat\mu_W-\mu_W)\rightarrow_d N(0, \sigma^2_W)$, where $\sigma^2_W=E \left\{ \left( \int _{0} ^{\tau} \tau ^{-1} \phi _{1} (t) dt \right)^{2} \right\}$ with $\phi_1(t)$ defined in equation (S14) of Web Appendix A. }

Detailed proofs of Theorems 1--2 are provided in Web Appendix A of the Supplementary Materials. 

By the proof of Theorem 2, we derive a closed form for the asymptotic variance of $\widehat\mu_W$. While the result naturally suggests a plug-in estimator of the variance of $\widehat\mu_W$, the analytic form of such a variance estimator depends  on  the form of {$\xi(\bz, t, \bO_i)$} in condition (C3), and thus the choice of the model for $C_i$ given $\cH_i(t)$. Alternatively, the standard nonparametric bootstrapping, which is readily justifiable by empirical process theory \citep{austin2022bootstrap, kosorok2008introduction, vaart2023empirical}, can provide a unified way to conduct variance estimation and other inferences, such as confidence intervals and hypothesis testing. Therefore, we recommend  adopting bootstrapping based inference procedures, which are outlined below.

Specifically, to estimate the variance of $\widehat\mu_W$, we first resample the observed data with replacement, and compute $\widehat\mu_W$ based on the bootstrapped sample, which is denoted by $\mu_W ^{*}$. Repeating  this procedure for $B$ times, where $B$ is a  pre-determined large number, we can obtain $B$ realizations of $\mu_W ^{*}$. Then we estimate the  variance of $\hat\mu_W$  by the empirical variance of $\mu _W ^{*}$. The confidence interval of $\mu_W$ can be obtained by using the empirical percentiles of $\mu_W^{*}$ or based on normal approximation. 

It is often of practical importance to compare the mean TIR between/among groups. To address such a problem, we can first obtain the proposed estimator of $\mu_W$ separately for each group, denoted by $\widehat \mu_{W, k}$, where $k=1,\ldots, K$ and $K$ represents the number of groups. Write $\widehat\bD_{\mu_W}=(\widehat\mu_{W, 2}-\widehat\mu_{W, 1}, \ldots, \widehat\mu_{W, K}-\widehat\mu_{W, 1})^\top$. A Wald-type test for the null hypothesis 
$H_0:\ \mu_{W,1}=\ldots=\mu_{W, K}$ can be constructed as
$
\widehat\bD_{\mu_W}^\top \widehat\bSigma_{D}^{-1} \widehat\bD_{\mu_W},
$
where $\widehat\bSigma_D$ is the covariance estimator of $\widehat\bD_{\mu_W}$ obtained by bootstrapping. We can obtain the $p$ value by comparing this test statistic with the $\chi_{K-1}^2$ distribution.

\section{Simulation Studies}
\label{sec:simulation}
We conduct extensive simulation studies to evaluate the finite sample performance of the proposed method and to illustrate its empirical advantages.
\subsection{Data generation and simulation set-ups}
\begin{figure}
    \centering
    \includegraphics[width = \textwidth]{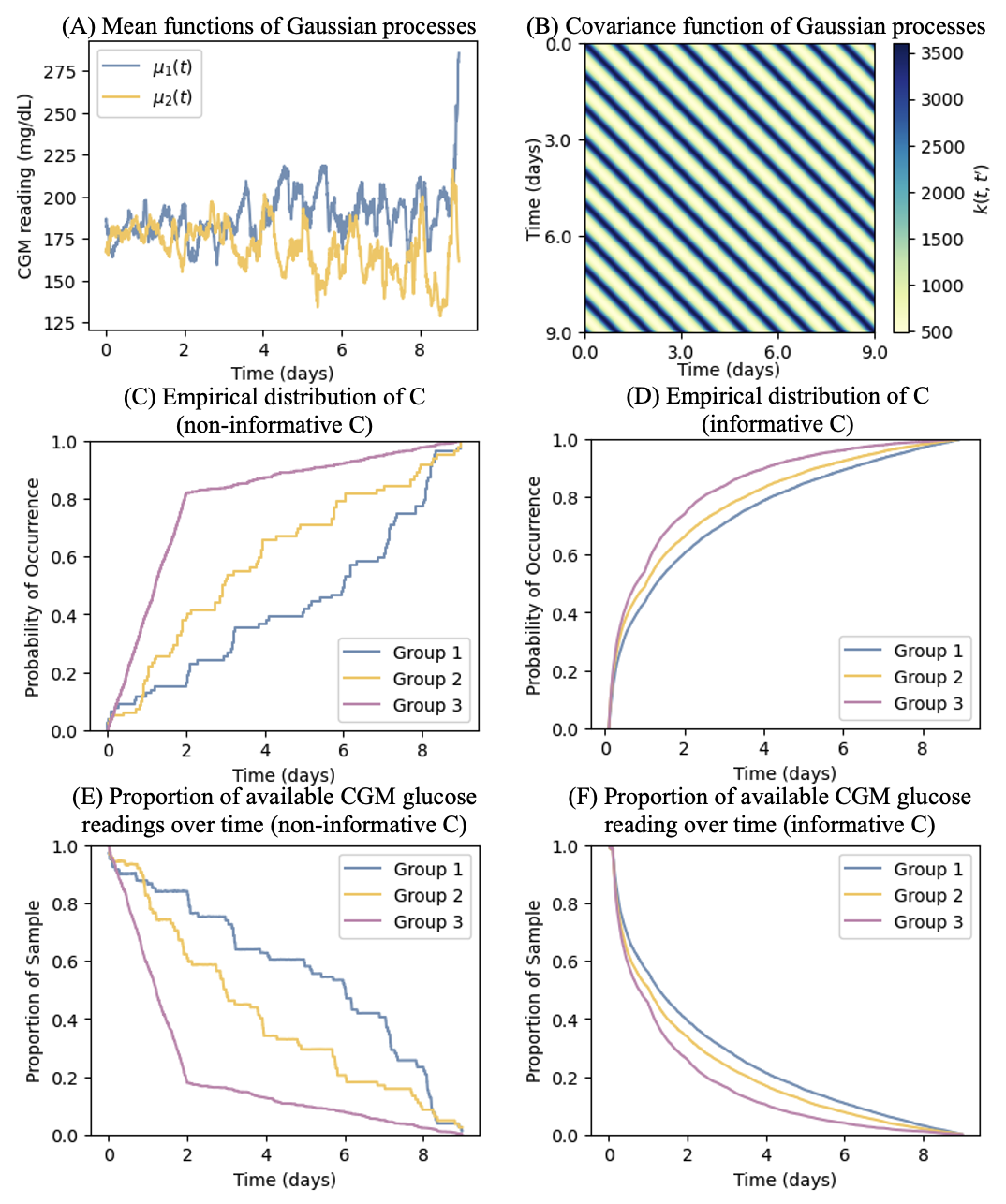}
    \caption{Configurations and characteristics of the simulated glucose trajectories, including (A): mean functions of  Gaussian processes; (B) covariance function of Gaussian processes; (C) empirical distribution of CGM follow-up duration $C$ when $C$ is non-informative; (D) empirical distribution of CGM follow-up duration $C$ when $C$ is informative; (E) proportion of available CGM glucose readings over time when the CGM follow-up duration $C$ is  non-informative; (F) proportion of available CGM glucose readings over time when the CGM follow-up duration $C$ is informative.}
    \label{fig:1}
\end{figure}

Our data generation scheme closely mimic real data scenarios in inpatient CGM studies. Specifically, we generate CGM glucose trajectories as Gaussian processes,  $\mathcal{GP}(\mu(t), k(t, t'))$, where $\mu(t)$ denotes the mean function and $k(t, t')$ denotes the covariance function characterized by a periodic kernel that takes the form, 
$ k(t, t') = \sigma ^{2} \text{exp} \left( - \frac{2}{l^2} \text{sin} ^2 \left( \pi \frac{|t - t'|}{p} \right) \right)$. The kernel parameters, amplitude (\(\sigma\)), length-scale (\(l\)), and period (\(p\)), govern the mean deviation, trajectory smoothness, and repetition frequency, respectively. To reflect the natural 1-day (i.e., 1440 minutes) periodicity in glucose, we  set the the  period \( p = 1440 \), length-scale \( l = 1 \), and amplitude \( \sigma = 62 \). The resulting covariance matrix is shown in Figure \ref{fig:1}(B).  

In each simulation setting, three groups of CGM glucose trajectories are generated and are referred to as Group 1, Group 2, and Group 3. For Groups 1 and 2, the corresponding mean functions, denoted by  $\mu_1(t)$ and $\mu_2(t)$, are specified as the empirical mean functions of the CGM glucose trajectories observed for the two study groups considered in the real data application presented in Section \ref{sec:realdata}. For Group 3, we use the same mean function specified for Group 1 so that comparing Group 1 versus Group 3 can serve as the null case with respective to the hypothesis testing on the equivalence in mean TIR. 
In addition, we set ${\mathcal T}$ as an equally spaced time grid with the step size of 5 minutes, mimicking the data capturing scheme of a Dexcom G6 CGM system. 

To induce missing data for the generated CGM glucose trajectories, we first impose intermittent missing. Specifically, we  draw the starting time of the intermittent missing from an exponential distribution with the scale parameter of 3424, which is derived from the real data application discussed in Section \ref{sec:realdata}, and then set the missing duration as a random number generated from a uniform distribution between 10 and 70 (minutes). By doing so, we determine the time interval when the intermittent missing occurs, and subsequently $\delta_i^*(t)$. 

Next, we generate $C_i$, which represents the ending time of CGM, in the following two ways: \\
(I) $C_i$ is completely {\it non-informative} and is sampled from a distribution not related to the generation of glucose trajectories. Specifically, for Groups 1 and 2,  we sample $C_i$'s with replacement from the empirical distribution of the observed CGM follow-up durations in the real data application discussed in Section \ref{sec:realdata}. For Group 3, $C_i$'s are generated from a mixture distribution, \(0.8F_1(c) + 0.2F_2(c) \), where \( F_1(\cdot) \) is the uniform distribution function between  0 to 2 (days) and \( F_2(\cdot) \) is the uniform distribution function between 2 and 9 (days). 
\\
(II) $C_i$ is {\it informative}, depending on the history of glucose trajectories, and is generated from the Cox's  proportional hazards model \eqref{eq:cox} with $\bZ_i(t)=(Z_i^{(1)}(t),\ Z^{(2)}_i)^\top$, where \(Z^{(1)}_i(t)\) captures the average glucose during the previous day as of time $t$ (which equals $0$ when $t$ indicates a time point during the first day), and \(Z^{(2)}_i\) is drawn from a uniform distribution \( \text{Unif}(-0.5, 0.5) \).  To make $Z^{(1)}_i(t)$ and $Z^{(2)}_i$ to range within a similar scale, we  normalize $Z^{(1)}_i(t)$ by dividing it by $100$. 
We let {$\lambda_0(t)= \frac{1}{16} t^{-15/16}$} and $\bbeta_0=(-2, -2)^\top$ for Group 1,  {$\lambda_0(t)= \frac{1}{4}t^{-3/4}$} and $\bbeta_0=(2, 2)^\top$ for Group 2, and {$\lambda_0(t)= \frac{1}{2} t^{-1/2}$} and $\bbeta_0=(2, 2)^\top$ for Group 3.

With the $\delta^*(t)$ and $C$ obtained as above, we compute the availability indicator $\delta(t)=\delta^*(t)I(C\leq t)$, and $Y(t)$ is observed if and only if $\delta(t)=1$.  The empirical distribution of $C$ and the empirical mean of $\delta(t)=1$, which reflects the proportion of the available glucose readings at time $t$, in case (I) and case (II)  are plotted in Figure \ref{fig:1}(C)--Figure \ref{fig:1}(F). 

In each simulation setting, we consider sample sizes $n=50, 100$ and $200$, and generate 1000 replicated datasets.  The bootstrap based inferences are implemented based on 200 bootstrapping samples.

\subsection{Simulation results}

In each simulation setting, we analyze the mean TIR over 7 days with the target glucose ranges, $<70$ mg/dL, between 70 and 180 mg/dL, and $>180$ mg/dL, using three different methods: (1) the oracle approach that computes $n^{-1}\sum_{i=1}^n \left(\int_0^\tau I(Y_i(t)\in G)dt\right)/\tau$ based on  the fully observed \(Y_i(\cdot)\) without any missing data;  (2) the proposed approach that uses \(\widehat{\mu} _W\);  (3) the naive approach that completely ignores the missing data and estimates \(\mu_W\) by \(\widecheck{\mu}_{W}\). 

Table \ref{tab:simulation} summarizes the simulation results in the setting with informative CGM follow-up durations (i.e., $C_i$'s). The presented results include the empirical averages, relative biases, empirical standard deviations, and estimated standard errors (based on bootstrapping) of the three different estimators of mean TIR. The relative bias is defined as the absolute difference between the average estimate and the average oracle estimate divided by the average oracle estimate. We also report the empirical rejection rates from testing the mean TIR equivalence between Groups 1 and 2 and between Groups 1 and 3. Since the mean TIRs are the same between Groups 1 and 3 and are different between Groups 1 and 2,  the empirical rejection rates in these cases respectively reflect  empirical size and empirical power. 

From Table \ref{tab:simulation}, we observe that the proposed estimator of mean TIR is virtually unbiased with the relative biases mostly  below 5\% and exhibiting a decreasing trend with the sample size. The empirical standard deviations and the estimated standard errors closely agree with each other.  The naive estimator demonstrates apparent biases in many cases. For example, even when the sample size is as large as 200, the relative biases of the naive estimates for the TIR $<70$ mg/dL are still pretty substantial, equal to 23.7\%, 18.7\%, and 33\% for Groups 1, 2, and 3 respectively. In contrast, the corresponding relative biases of the proposed estimator are much smaller, equal to 0.4\%, 0.1\%, and 6.6\%. It is noted that the proposed estimator generally produces larger relative biases for Group 3  as compared to that for Group 1, though Group 1 and Group 3 share the same mean TIR. This can be explained by the different distributions of $C$ between these two groups (see Figure \ref{fig:1}(D)), which entail more frequent early termination of CGM and thus more intensive monotone missing in the data for Group 3 (as confirmed by Figure \ref{fig:1}(F)).  As expected, the lower proportions of available CGM glucose readings in Group 3 (as compared to Group 1)  also result in larger estimation variability captured by either empirical standard deviations or estimated standard errors. 

Regarding the hypothesis testing that compares the mean TIR between Groups 1 and 3, the proposed method yields  empirical sizes  quite close to the nominal significance level of 5\%, while the naive approach may be prone to inflated type I errors (e.g.,  the empirical size of 10\% when comparing the mean TIR $>180$ mg/dL with $n=200$). By examining the empirical rejection rates for comparing Groups 1 vs. 2, we find that the test based on the proposed method is much more powerful  as compared to the test based on the naive approach, as evidenced by more than two fold larger empirical rejection rates. For example, with the larger sample size 200, the naive method only has around 11\% power to detect the difference in mean TIR between Group 1 and Group 2, while the proposed method has 74\% power to do so. 

The simulation results in the setting with non-informative $C_i$'s are presented in Table S1 of Web Appendix B.1. The finding from Table S1 is consistent with that  from Table \ref{tab:simulation}. Both suggest that failing to properly account for the special missing features of the inpatient CGM glucose trajectories  can lead to considerably biased estimation of mean TIR as well as problematic inferences that are reflected by the inflated type I errors and substantial power loss associated with the naive method.

\begin{table}
    \centering
    \caption{Simulation results in the case with informative CGM follow-up durations, including empirical averages (AvgEst), relative biases (RelBias), empirical standard deviations (ESD), and estimated standard errors (BSE)  of the oracle, naive, proposed estimators  of mean TIR, and the empirical rejection rates from testing the mean TIR equivalence  between Group 1 and Group 3 (Size) and between Group 1 and Group 2 (Power).}
    \resizebox{\textwidth}{!}{
    \begin{tabular}{ccccccccccc}
\hline
Sample size & & \multicolumn{9}{c}{Target glucose range}\\
per group   & & \multicolumn{3}{c}{$< 70mg/dL$} & \multicolumn{3}{c}{Between $70$ and $180mg/dL$} & \multicolumn{3}{c}{$>180mg/dL$}\\
 n = 50   &         & Oracle   & Naive   & Proposed   & Oracle   & Naive   & Proposed   & Oracle   & Naive   & Proposed   \\
\hline
 Group 1  & AvgEst (\%)          & 1.18    & 1.45   & 1.2       & 43.26   & 47.18   & 43.4       & 55.26    & 50.97   & 55.1       \\
          & RelBias (\%)      & -        & 23.3    & 1.4        & -        & 9.1     & 0.3        & -        & 7.8     & 0.3        \\
          & BSE ($\times 10^2$)            & 0.51     & 0.64    & 0.66       & 3.74     & 3.94    & 5.64       & 3.94     & 4.14    & 5.89       \\
          & ESD ($\times 10^2$)            & 0.51     & 0.64    & 0.66       & 3.73     & 3.93    & 5.62       & 3.93     & 4.13    & 5.87        \\
 Group 2  & AvgEst (\%)          & 2.56     & 2.08    & 2.61       & 51.74    & 49.28   & 51.85      & 44.72    & 47.91   & 44.51      \\
          & RelBias (\%)      & -        & 18.6    & 1.9        & -        & 4.8     & 0.2        & -        & 7.1     & 0.5        \\
          & BSE ($\times 10^2$)            & 0.78     & 0.78    & 1.25       & 3.63     & 3.91    & 5.81       & 3.94     & 4.18    & 6.38      \\
          & ESD ($\times 10^2$)            & 0.78     & 0.77    & 1.24       & 3.62     & 3.9     & 5.79       & 3.93     & 4.17    & 6.36        \\
 Group 3  & AvgEst (\%)          & 1.17     & 1.56    & 1.4        & 43.13    & 48.52   & 45.57      & 55.39    & 49.47   & 52.63      \\
          & RelBias (\%)      & -        & 33.7    & 19.4       & -        & 12.5    & 5.7        & -        & 10.7    & 5.0        \\
          & BSE ($\times 10^2$)            & 0.51     & 0.71    & 0.93       & 3.77     & 4.09    & 7.88       & 3.95     & 4.28    & 8.32       \\
          & ESD ($\times 10^2$)            & 0.51     & 0.7     & 0.91       & 3.76     & 4.08    & 7.52       & 3.94     & 4.27    & 7.95       \\
 Power    & Groups 1 vs 2 & 29.4\%    & 8.2\%    & 21.3\%      & 36.6\%    & 6.9\%    & 31.6\%      & 46.7\%    & 9.1\%    & 34.4\%      \\
 Size     & Groups 1 vs 3 & 4.7\%     & 6.8\%    & 4.4\%       & 5.5\%     & 9.2\%    & 6.7\%      & 5.0\%     & 9.3\%    & 6.7\%      \\
\hline
 n = 100   &         &    &    &    &    &    &    &    &    &    \\
\hline
 Group 1   & AvgEst (\%)          & 1.18     & 1.47    & 1.16       & 43.33    & 47.26   & 43.24      & 55.16    & 50.85   & 55.26      \\
           & RelBias (\%)      & -        & 24.6    & 1.3        & -        & 9.1     & 0.2        & -        & 7.8     & 0.2        \\
           & BSE ($\times 10^2$)            & 0.37     & 0.47    & 0.5        & 2.67     & 2.81    & 4.03       & 2.81     & 2.96    & 4.22        \\
           & ESD($\times 10^2$)            & 0.37     & 0.47    & 0.5        & 2.66     & 2.8     & 4.02       & 2.8      & 2.95    & 4.21        \\
 Group 2   & AvgEst (\%)          & 2.59     & 2.1     & 2.58       & 51.72    & 49.35   & 51.63      & 44.69    & 47.79   & 44.77      \\
           & RelBias (\%)      & -        & 18.7    & 0.2        & -        & 4.6     & 0.2        & -        & 7.0     & 0.2        \\
           & BSE ($\times 10^2$)            & 0.56     & 0.57    & 0.94       & 2.58     & 2.79    & 4.23       & 2.8      & 2.98    & 4.6        \\
           & ESD ($\times 10^2$)            & 0.56     & 0.56    & 0.94       & 2.57     & 2.78    & 4.22       & 2.79     & 2.97    & 4.59       \\
 Group 3   & AvgEst (\%)          & 1.18     & 1.58    & 1.35       & 43.4     & 48.79   & 44.56      & 55.1     & 49.19   & 53.73      \\
           & RelBias (\%)      & -        & 33.2    & 14.4       & -        & 12.4    & 2.7        & -        & 10.7    & 2.5        \\
           & BSE ($\times 10^2$)            & 0.37     & 0.52    & 0.78       & 2.68     & 2.91    & 6.89       & 2.81     & 3.05    & 7.21        \\
           & ESD ($\times 10^2$)            & 0.37     & 0.52    & 0.76       & 2.67     & 2.91    & 6.58       & 2.8      & 3.05    & 6.88       \\
 Power     & Groups 1 vs 2 & 55.8\%    & 14.4\%   & 31.5\%      & 60.2\%    & 9.8\%    & 52.2\%      & 72.7\%    & 12.7\%   & 59.3\%      \\
 Size      & Groups 1 vs 3 & 4.8\%     & 7.2\%    & 6.6\%       & 6.3\%     & 9.7\%    & 7.7\%      & 6.4\%     & 9.6\%    & 6.7\%      \\
\hline
 n = 200   &         &    &    &    &    &    &    &    &    &    \\
\hline
 Group 1   & AvgEst (\%)          & 1.19     & 1.48    & 1.19       & 43.39    & 47.32   & 43.38      & 55.09    & 50.78   & 55.1       \\
           & RelBias (\%)      & -        & 23.7    & 0.3        & -        & 9.1     & 0.0        & -        & 7.8     & 0.0        \\
           & BSE ($\times 10^2$)            & 0.27     & 0.34    & 0.38       & 1.89     & 1.99    & 2.89       & 1.99     & 2.09    & 3.02       \\
           & ESD ($\times 10^2$)            & 0.27     & 0.34    & 0.37       & 1.88     & 1.98    & 2.88       & 1.98     & 2.08    & 3.01       \\
 Group 2   & AvgEst (\%)          & 2.6      & 2.11    & 2.6        & 51.69    & 49.31   & 51.52      & 44.7     & 47.83   & 44.88      \\
           & RelBias (\%)      & -        & 18.7    & 0.1        & -        & 4.6     & 0.3        & -        & 7.0     & 0.4        \\
           & BSE ($\times 10^2$)            & 0.4      & 0.41    & 0.7        & 1.83     & 1.97    & 3.05       & 2.0      & 2.12    & 3.34       \\
           & ESD ($\times 10^2$)            & 0.4      & 0.41    & 0.7        & 1.83     & 1.97    & 3.04       & 1.99     & 2.11    & 3.33       \\
 Group 3   & AvgEst (\%)          & 1.18     & 1.57    & 1.26       & 43.41    & 48.85   & 44.15      & 55.08    & 49.12   & 54.25      \\
           & RelBias (\%)      & -        & 33.0    & 6.6        & -        & 12.5    & 1.7        & -        & 10.8    & 1.5        \\
           & BSE ($\times 10^2$)            & 0.27     & 0.37    & 0.62       & 1.89     & 2.05    & 5.83       & 1.98     & 2.16    & 6.14       \\
           & ESD ($\times 10^2$)            & 0.27     & 0.37    & 0.61       & 1.89     & 2.05    & 5.61       & 1.98     & 2.15    & 5.9       \\
 Power     & Groups 1 vs 2 & 84.4\%    & 21.1\%   & 71.3\%      & 88.4\%    & 10.6\%   & 73.7\%      & 96.0\%    & 16.0\%   & 82.1\%      \\
 Size      & Groups 1 vs 3 & 5.3\%     & 7.0\%    & 6.2\%      & 6.7\%     & 9.9\%    & 7.0\%      & 6.5\%     & 10.0\%    & 6.8\%      \\
\hline
\end{tabular}
}
    \label{tab:simulation}
\end{table}

\subsection{Simulation studies under different configurations}

We conduct simulation studies under other configurations that represent different realistic characteristics of inpatient CGM data. For example, we consider settings with different sizes of group difference by modulating the mean functions used to generate CGM glucose trajectories.   We also simulate data with  different missing data patterns by varying the distributions of $C$. Details on these additional simulation studies are provided in Web Appendix B.2. The results consistently demonstrate the advantages of the proposed method  over the naive method. 

\subsection{Simulation studies on sensitivity analysis}

We further conduct simulation studies to evaluate the robustness of the proposed method to the misspecification of the assumed Cox model for $C$ given $\cH(t)$. Specifically, we generate the CGM glucose trajectory $Y_i(\cdot)$ as a Gaussian process with the mean function $\mu^{[i]}(t)$ and the covariance function $k(t, t')$, which is the same as that  plotted in Figure \ref{fig:1}(B). We set the mean function $\mu^{[i]}(t)=\mu_1(t)-3\zeta_i$ for Group 1 or 3 and $\mu^{[i]}(t)=\mu_2(t)+3\zeta_i $ for Group 2, where $\mu_1(t)$ and $\mu_2(t)$ are the same as those plotted in Figure \ref{fig:1}(A) and $\zeta_i$ is a $Bernoulli (0.5)$ random variable. We generate the ending time of CGM, $C_i$, based on a transformation model \citep{cheng1995analysis}, $C_i=s\cdot\exp(-\zeta_i+\epsilon_i)$, where $s=8$ for Group 1 and Group 2 and $s=3$ for Group 3, and the error term $\epsilon_i$  has the distribution function, $p F_3(x)+(1-p) F_4(x)$. Here $F_3(\cdot)$ denotes the standard logistic distribution function and $F_4(\cdot)$ denotes the extreme value distribution function. It is clear that $C_i$ follows the Cox proportional hazards model \eqref{eq:cox} with $\bZ_i(t)=\zeta_i$ when $p=0$ and a proportional odds model when $p=1$. As such, the magnitude of the parameter $p$ reflects the degree of departure from a Cox model. 

Table \ref{tab:simulation_sensitivity} presents the simulation results for the settings described above with $p=0.5$ and $p=1$. It is shown that the proposed method yields quite small relative biases, ranging from 1.2\% to 2.0\%, when $p=0.5$ or $1$. The empirical sizes for testing the mean TIR equivalence between Group 1 and Group 3 are still quite close to the nominal significance level of $5\%$. At the same time, the relative biases of the naive estimator are several fold larger than those of the proposed estimator. The naive method also results in tests which yield inflated type-I errors or suffer from substantial power loss.  These observations suggest that  the proposed method can offer robust estimation and inference on mean TIR and consistently outperform the naive method even when the Cox model assumption for  the CGM follow up duration (i.e., $C_i$) does not hold. 

\begin{table}[H]
    \centering
    \caption{Simulation results from sensitivity analyses including empirical averages (AvgEst), relative biases (RelBias), empirical standard deviations (ESD), and estimated standard errors (BSE) of the oracle, naive, proposed estimators of mean TIR, and the empirical rejection rates from testing the mean TIR equivalence between Group 1 and Group 3 (Size) and between Group 1 and Group 2 (Power).}
    \resizebox{\textwidth}{!}{
    \begin{tabular}{ccccccccccc}
\hline
Degree of departure & Sample size & \multicolumn{9}{c}{Target glucose range}\\
from a Cox model   & per group & \multicolumn{3}{c}{$< 70mg/dl$} & \multicolumn{3}{c}{Between $70$ and $180mg/dl$} & \multicolumn{3}{c}{$>180mg/dl$}\\
\hline
 p = 0.5   & n = 200  & Oracle   & Naive   & Proposed   & Oracle   & Naive   & Proposed   & Oracle   & Naive   & Proposed   \\
\hline
 Group 1   & AvgEst (\%)          & 1.26     & 1.5     & 1.24       & 44.47    & 47.51   & 43.76      & 53.92    & 50.56   & 54.73      \\
           & RelBias (\%)       & -        & 19.1    & 1.9        & -        & 6.8     & 1.6        & -        & 6.2     & 1.5        \\
           & BSE ($\times 10^2$)              & 0.28     & 0.37    & 0.33       & 1.89     & 2.07    & 2.39       & 1.99     & 2.16    & 2.51        \\
           & ESD ($\times 10^2$)             & 0.28     & 0.37    & 0.33       & 1.9      & 2.08    & 2.4        & 1.99     & 2.17    & 2.51      \\
 Group 2   & AvgEst (\%)          & 2.43     & 2.15    & 2.49       & 50.85    & 49.39   & 51.56      & 45.8     & 47.7    & 45.07      \\
           & RelBias (\%)       & -        & 11.7    & 2.7        & -        & 2.9     & 1.4        & -        & 4.1     & 1.6        \\
           & BSE ($\times 10^2$)              & 0.39     & 0.43    & 0.51       & 1.84     & 2.03    & 2.32       & 2.0      & 2.17    & 2.51       \\
           & ESD ($\times 10^2$)              & 0.39     & 0.43    & 0.51       & 1.84     & 2.04    & 2.33       & 2.0      & 2.17    & 2.52       \\
 Group 3   & AvgEst (\%)          & 1.27     & 1.64    & 1.24       & 44.29    & 48.83   & 43.4      & 54.08    & 49.05   & 55.11      \\
           & RelBias (\%)       & -        & 28.9    & 2.4        & -        & 10.2    & 2.0        & -        & 9.3     & 1.9        \\
           & BSE ($\times 10^2$)             & 0.28     & 0.43    & 0.43       & 1.89     & 2.24    & 3.18       & 1.99     & 2.34    & 3.34       \\
           & ESD ($\times 10^2$)               & 0.28     & 0.44    & 0.43       & 1.9      & 2.25    & 3.19       & 2.0      & 2.34    & 3.35       \\
 Power     & Groups 1 vs 2 & 69.4\%    & 22.2\%   & 57.3\%      & 68.8\%    & 9.6\%    & 59.6\%      & 83.0\%    & 14.6\%   & 74.5\%      \\
 Size      & Groups 1 vs 3 & 4.6\%     & 7.8\%    & 5.3\%       & 6.0\%     & 10.3\%    & 4.8\%       & 5.4\%     & 9.7\%    & 4.6\%       \\
\hline
 p = 1   & n = 200  & Oracle   & Naive   & Proposed   & Oracle   & Naive   & Proposed   & Oracle   & Naive   & Proposed   \\
\hline
 Group 1   & AvgEst (\%)          & 1.27     & 1.49    & 1.26       & 44.35    & 47.07   & 43.7       & 54.01    & 51.01   & 54.79      \\
           & RelBias (\%)      & -        & 17.0    & 1.2        & -        & 6.1     & 1.5        & -        & 5.6     & 1.5        \\
           & BSE ($\times 10^2$)            & 0.28     & 0.36    & 0.33       & 1.9      & 2.07    & 2.31       & 2.0      & 2.16    & 2.44       \\
           & ESD ($\times 10^2$)            & 0.28     & 0.36    & 0.33       & 1.89     & 2.06    & 2.31       & 1.99     & 2.16    & 2.44       \\
 Group 2   & AvgEst (\%)          & 2.47     & 2.22    & 2.51       & 50.98    & 49.77   & 51.67      & 45.6     & 47.2    & 44.89      \\
           & RelBias (\%)      & -        & 10.1    & 1.7        & -        & 2.4     & 1.4        & -        & 3.5     & 1.6        \\
           & BSE ($\times 10^2$)            & 0.39     & 0.43    & 0.49       & 1.85     & 2.03    & 2.24       & 2.0      & 2.17    & 2.45       \\
           & ESD ($\times 10^2$)            & 0.39     & 0.43    & 0.49       & 1.84     & 2.03    & 2.23       & 2.0      & 2.17    & 2.44       \\
 Group 3   & AvgEst (\%)          & 1.26     & 1.58    & 1.22       & 44.28    & 48.37   & 43.46      & 54.11    & 49.59   & 55.06      \\
           & RelBias (\%)      & -        & 25.9    & 3.3        & -        & 9.2     & 1.9        & -        & 8.3     & 1.8        \\
           & BSE ($\times 10^2$)            & 0.28     & 0.43    & 0.38       & 1.89     & 2.23    & 2.84       & 1.99     & 2.33    & 2.98       \\
           & ESD ($\times 10^2$)            & 0.28     & 0.42    & 0.38       & 1.89     & 2.23    & 2.83       & 1.99     & 2.32    & 2.97       \\
 Power     & Groups 1 vs 2 & 70.4\%    & 27.2\%   & 55.6\%      & 71.7\%    & 15.7\%   & 60.7\%      & 85.1\%    & 23.1\%   & 74.7\%      \\
 Size      & Groups 1 vs 3 & 4.5\%     & 6.2\%    & 5.4\%       & 6.1\%     & 8.8\%    & 5.3\%       & 5.3\%     & 9.2\%    & 5.8\%       \\
\hline
\end{tabular}
}
    \label{tab:simulation_sensitivity}
\end{table}

\section{A Real Data Application}
\label{sec:realdata}

We apply the proposed method to a combined dataset from two inpatient CGM studies, Dexcom G6 Observational Study \citep{davis2021accuracy} and Dexcom G6 Interventional Study \citep{spanakis2022continuous}.  Dexcom G6 Observational Study was an observational study of 91 insulin-treated adult medicine and surgery patients with type 2 diabetes, aiming to evaluate the feasibility of using the Dexcom G6 CGM in hospitals. Dexcom G6 Interventional Study was a randomized multi-center clinical trial which was aimed to assess the efficacy of Dexcom G6 CGM in guiding insulin adjustment in insulin-treated adult medicine and surgery patients with diabetes. 
Among  the 185 patients who consented to participate in this study,  91 patients were randomized to receive the standard of care  with  insulin dose adjusted based on capillary point-of care glucose monitoring while wearing a blinded Dexcom G6 CGM (i.e., control group), and 94 patients were randomized to receive insulin adjustment based on daily CGM profile (i.e., intervention group). 
The Dexcom G6 Observational Study and the Dexcom G6 Interventional Study adopted the same inclusion and exclusion criteria for study enrollment and the same standard of care protocol, which was implemented for all participants of Dexcom G6 Observational Study and the control group of Dexcom G6 Interventional Study.  After pooling the datasets from these two inpatient CGM studies, we use the combined dataset to evaluate the effect of CGM-guided insulin adjustment on  various TIR outcomes.

In our analyses, the TIR outcomes of interest include TIR defined for six glycemic ranges, $<70$ mg/dL, between $70$ and $180$ mg/dL,  between $70$ and $140$ mg/dL, $>180$mg/dL, $>140$mg/dL, and $>250$mg/dL, and over time periods of $1$, $3$, $5$, $7$ and $9$ days. The ranges of 70--180 mg/dL and 70--140 mg/dL represent two commonly adopted glycemic control targets in hospitals with the latter one being more strict than the former one. The ranges of $>140$ mg/dL, $>180$ mg/dL, and $>250$mg/dL respectively correspond to borderline hyperglycemia, hyperglycemia, and severe hyperglycemia. The range of $<70$ mg/dL is considered to mark the time under mild hypoglycemia. 
After removing patients who had less than 24 hours of hospital stay, the final dataset includes 249 patients with 167 patients in the control group and 82 patients in the intervention group. 

The summary statistics with respect to age, gender, race, BMI, and obesity status are presented in Table S4 of Web Appendix B, showing similar patient characteristics between the intervention group and  the control group. From Table S4, we observe a trend of  longer CGM follow-up duration for the control group as compared to the intervention group (e.g., median CGM follow-up duration of 91.8 vs. 71.0 hours). The shorter CGM follow-up durations, which implicate shorter lengths of hospital stay,  in the intervention group may be related to better glycemic control resulted from the CGM-guided intervention;  hence $C_i$'s  in this dataset are likely informative about the underlying glycemic control. The proposed method can properly account for such informative $C_i$'s  by allowing $C_i$ to depend on CGM glucose history. 

\begin{figure}
    \centering
    \includegraphics[width=\linewidth]{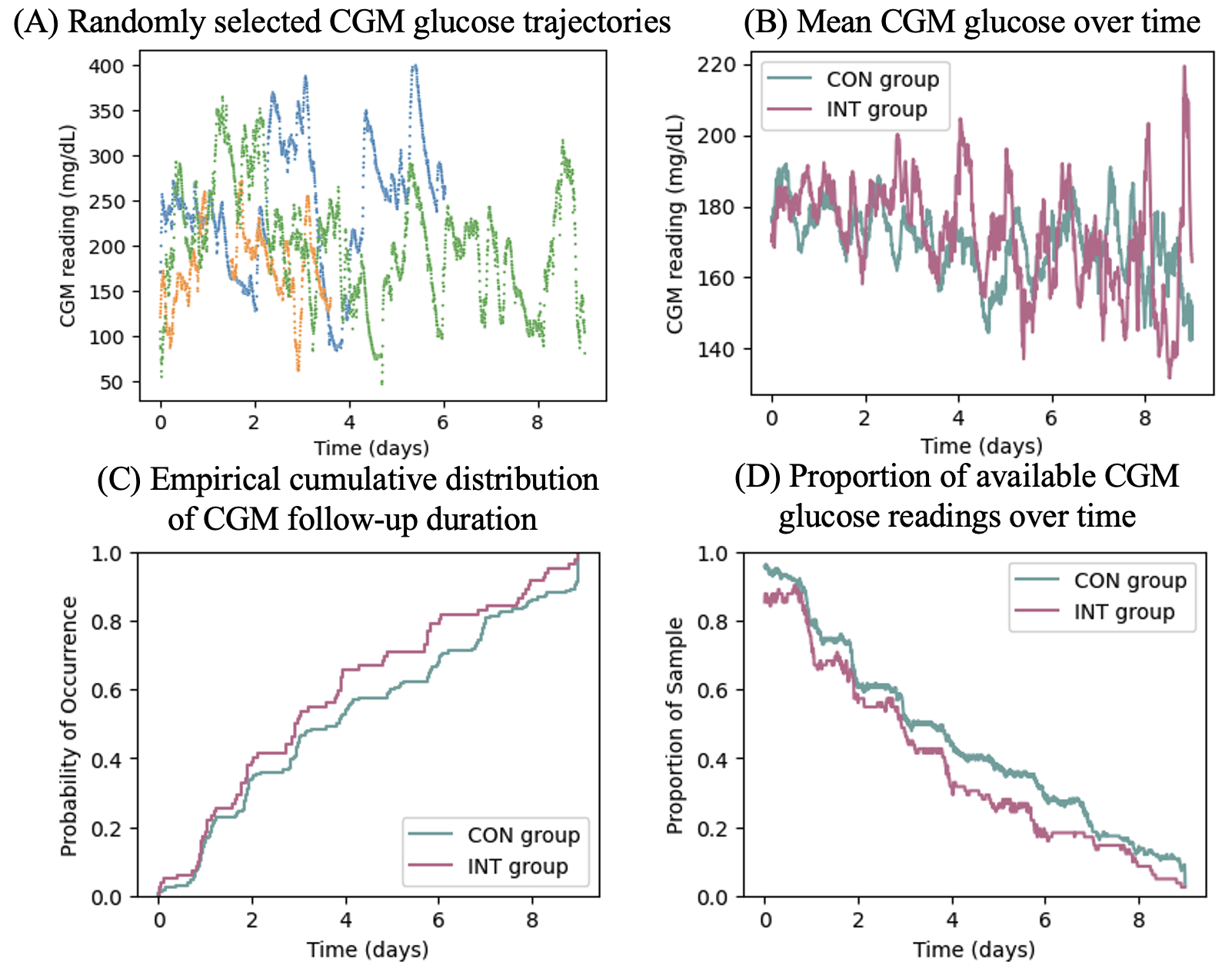}
    \caption{Illustration and summaries of CGM glucose data for the control (CON) and intervention (INT) groups including: (A) three randomly selected CGM glucose trajectories; (B) empirical average of available CGM glucose readings over time; (C) empirical cumulative distribution function of CGM follow-up duration; (D) proportion of available CGM glucose readings over time.}
    \label{fig:2}
\end{figure}

In Figure \ref{fig:2}, we illustrate and summarize the CGM glucose trajectories in our dataset. Figure \ref{fig:2}(A) presents three randomly selected CGM glucose trajectories. These trajectories shows the heterogeneous temporal patterns  across individuals as well as the subject-varying follow-up duration of CGM.   Figure \ref{fig:2}(B) presents the average  CGM glucose readings over time for  the control group and the interventional group, which entail the overall non-stationary patterns of inpatient CGM glucose trajectories.  In Figure \ref{fig:2}(C), we plot the empirical distributions of CGM follow-up duration $C_i$ separately for the control group and the intervention group, suggesting a trend of shorter CGM follow-up durations in the intervention group which result in lower proportions of available CGM glucose reading over time, as shown in Figure \ref{fig:2}(D). 

\begin{figure}
    \centering
    \includegraphics[width = \textwidth]{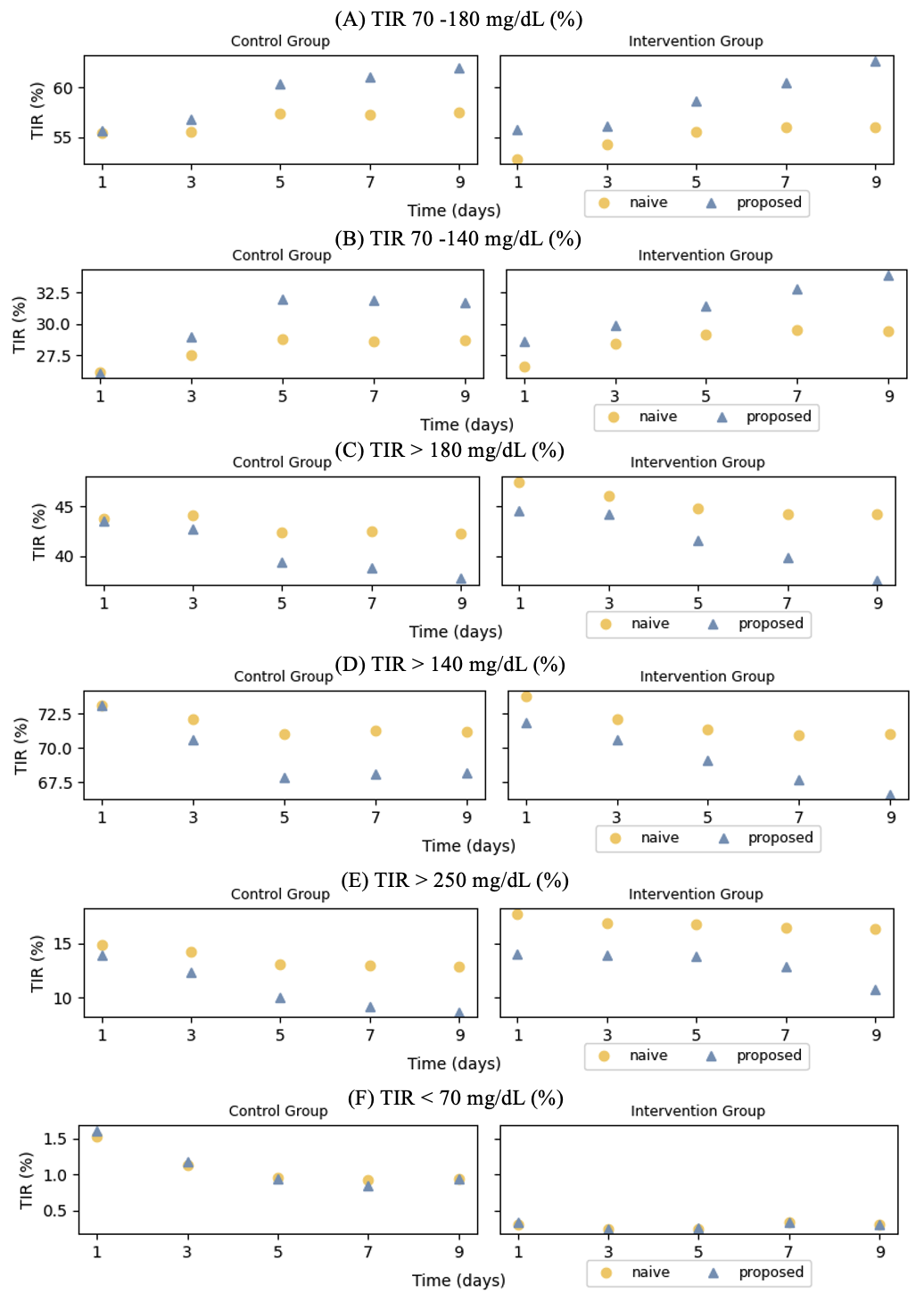}
    \caption{Estimates for mean TIR 70 - 180mg/dL, TIR 70 - 140 mg/dL, TIR $>$ 180 mg/dL, TAR $>$ 140 mg/dL, TIR $>$ 250 mg/dL, and TIR $<$ 70 mg/dL over the first 1, 3, 5, 7, and 9 days obtained by the proposed method and the naive method.}
    \label{fig:3}
\end{figure}

In Figure \ref{fig:3}, we present the proposed estimates and the naive estimates for the mean TIR outcomes described earlier, which correspond to $\mu_W$ with $G=[70, 180]$, $[70, 140]$, $(180, \infty)$, $(140, \infty)$, $(250, \infty)$, $(0, 70)$ and  $\tau=1, 3, 5, 7, 9$ (days), separately  for the control and intervention groups. We observe notable discrepancies between the proposed estimates and the naive estimates. For mean TIR 70-180 mg/dL and mean TIR 70-140 mg/dL, the proposed estimates are generally larger than the naive estimates. This is well expected because of a typical inpatient glucose pattern of being higher (and then more out-of-target glucose values) at the beginning of a hospital stay   and gradually stabilized over time. The naive method  takes the CGM glucose pattern observed for a hospital stay shorter than the required amount of time as the representative pattern for the whole time period. Doing so essentially overweighs the lower within-target proportions inherited with the initial CGM period  in hospital, and thus yields a mean TIR estimate biased downwards. The same explanation can be adapted to explain the lower estimated TIR by the proposed method (as compared to the naive estimate) with the glycemic ranges being above 140, 180, or 250 mg/dL.  As the patient cohorts of the Dexcom G6 Observation and Interventional Studies generally have small TIR $<70$ mg/dL, the differences between the proposed estimates and the naive estimates for the corresponding mean TIR have small magnitudes, as shown by Figure \ref{fig:3}.

We compare the  mean TIR outcomes  between the control group and the intervention group. In Table \ref{table:TIR_realdata}, we present the estimates and the corresponding standard errors for mean TIRs with $\tau=7$ (days) as well as the p values for testing the mean TIR equivalence between the two groups (which are obtained by bootstrapping). The proposed and naive methods attain similar conclusions regarding the comparisons of mean TIRs between the control and intervention groups.  
The results from both methods suggest similar mean TIRs for the glycemic targets of $70-180$ mg/dL and $70-140$ mg/dL, and the ranges of above $>180, 140$, and $250$ mg/dL between the control and intervention groups. At the same time,  applying the proposed method yields the estimated mean TIR for the intervention group equal to 0.34\%, which is much lower than that for the control group, which equals 0.84\%. The $p$ value for comparing the mean TIR between these two groups is $0.01$. This result  confirms the previous finding that CGM guided insulin adjustment may help reduce the percent time in hypoglycemia \citep{spanakis2022continuous}. Through properly handling the incompletely observed CGM glucose trajectories over the first 7 days which are commonly presented in our dataset, the proposed analysis evidences  the benefit of CGM in hypoglycemia prevention and correction with enhanced statistical rigor.

The proposed estimation and inference regarding the TIR outcomes over 1, 3, 5, 9 days yield consistent findings. The detailed results are provided in Tables S5-S8 of Web Appendix B. The consistent findings across the TIR outcomes defined with different CGM lengths consolidate the conclusions drawn from our analyses.

\begin{table}[H]
\caption{Results from analyzing the real CGM data including estimates and corresponding standard errors for mean TIRs over the first 7 days, and p values from testing the mean TIR equivalence between the control group and the intervention group by the naive method and the proposed method.}
\label{table:TIR_realdata}
\begin{threeparttable}
\resizebox{\textwidth}{!}{
\begin{tabular}{lccc}
\hline
                              & \multicolumn{2}{c}{Naive Method}                               & \multicolumn{1}{l}{} \\
                              & Control group (N = 167)             & Intervention group (N = 82)            & P value \tnote{a}              \\ 
\hline
 TIR 70 - 180 mg/dL (\%) & 57.33$\pm$2.14 & 56.05$\pm$3.48 & 0.75 \\
 TIR 70 - 140 mg/dL (\%) & 28.63$\pm$1.88 & 29.53$\pm$2.78 & 0.79 \\
 TIR \ensuremath{>} 180 mg/dL (\%)    & 42.45$\pm$2.15 & 44.17$\pm$3.31 & 0.66 \\
 TIR \ensuremath{>} 140 mg/dL (\%)    & 71.28$\pm$1.84 & 70.92$\pm$2.86 & 0.92 \\
 TIR \ensuremath{>} 250 mg/dL (\%)    & 12.97$\pm$1.4  & 16.48$\pm$2.67 & 0.24 \\
 TIR \ensuremath{<} 70 mg/dL (\%)    & 0.91$\pm$0.16  & 0.34$\pm$0.1   & \ensuremath{<}0.01    \\
\hline
                              & \multicolumn{2}{c}{Proposed Method}        &                      \\
\hline
 TIR 70 - 180 mg/dL (\%) & 61.06$\pm$2.12 & 60.43$\pm$3.92 & 0.89 \\
 TIR 70 - 140 mg/dL (\%) & 31.92$\pm$2.18 & 32.76$\pm$3.53 & 0.84 \\
 TIR \ensuremath{>} 180 mg/dL (\%)   & 38.81$\pm$2.23 & 39.81$\pm$3.84 & 0.82 \\
 TIR \ensuremath{>} 140 mg/dL (\%)    & 68.15$\pm$2.18 & 67.71$\pm$3.48 & 0.92 \\
 TIR \ensuremath{>} 250 mg/dL (\%)   & 9.16$\pm$0.99  & 12.83$\pm$2.45 & 0.17 \\
 TIR \ensuremath{<} 70 mg/dL (\%)    & 0.84$\pm$0.15  & 0.34$\pm$0.14  & 0.01 \\
\hline
\end{tabular}
}
\begin{tablenotes}
%\centering
\item[a] The p values are calculated using the Wald-type test described in Section 3.
\end{tablenotes}
\end{threeparttable}
\end{table}

\section{Remarks}
\label{sec:remarks}
This work represents a timely effort to develop rigorous statistical estimation and inference tools for the analysis of inpatient CGM studies which are on a rapid rise since the Covid-19 pandemic. Through formulating the real CGM glucose data as functional data subject to missing,  we develop viable and robust statistical strategies to deal with the special data features inherited with inpatient CGM studies, which however are commonly ignored in routine data analyses. Our proposals have solid theoretical underpinning while enjoying simple and stable implementation. The Python code for implementing the proposed method is publicly available on the GitHub repository: \url{https://github.com/qyxxx/CGM_meanTIR}. Future development may involve an R version. Such updates will be made available on the same GitHub repository. 

\backmatter

\section*{Acknowledgements}

The proposed research has been sponsored by the funding from the National Institutes of Health (Grant No.: R01DKDK136023).
\vspace*{-8pt}

\section*{Supplementary Materials}

Web Appendices, referenced in Section~\ref{sec:simulation} and Section~\ref{sec:realdata}, is available with
this paper at the Biometrics website on Wiley Online
Library.\vspace*{-8pt}

\bibliographystyle{biom} 
\bibliography{reference_peng}

\label{lastpage}

\end{document}